\documentclass[english,11pt,a4paper]{article}
\usepackage[T1]{fontenc}
\usepackage[latin9]{inputenc}
\usepackage{amsmath}
\usepackage{amsthm}
\usepackage{amssymb}
\usepackage{graphicx}

\makeatletter
\numberwithin{equation}{section}

\usepackage{jheppub}
\usepackage{orcidlink}
\usepackage{fontawesome5}
\allowdisplaybreaks

\DeclareDocumentCommand{\GitHub}{s}{%
	\IfBooleanTF{#1}{%
		\faGithub~\url{https://github.com/Fenyutanchan/Primordial-Black-Hole.git}%
	}{%
		\href{https://github.com/Fenyutanchan/Primordial-Black-Hole.git}{\faGithub}%
	}%
}

\makeatother

\usepackage{babel}
\begin{document}
\title{Primordial Black Holes Evaporating before Big Bang Nucleosynthesis}

\abstract{  Primordial black holes (PBHs) formed from the collapse
of density fluctuations provide a unique window into the physics of
the early Universe. Their evaporation through Hawking radiation around
the epoch of Big Bang nucleosynthesis (BBN) can leave measurable imprints
on the primordial light-element abundances.  In this work, we analyze
in detail the effects of PBHs evaporating before BBN, with various
intermediate steps understood analytically, and obtain the BBN constraint
on PBHs within a transparent and reproducible framework. We find that,
to produce observable effects on BBN, the PBH mass must exceed $10^{9}$
g, a threshold higher than that reported in some earlier studies.
Slightly above $10^{9}$ g, the BBN sensitivity rapidly increases
with the mass and then decreases, with the turning point occurring
 at $2\times10^{9}$ g.  For PBHs in the mass range $[10^{9},\ 10^{10}]$
g, current measurements of BBN observables set an upper bound on the
initial mass fraction parameter $\beta$ ranging from $10^{-17}$
to $10^{-19}$.  To facilitate future improvements, we make our code
publicly available, enabling straightforward incorporation of updated
nuclear reaction rates, particle-physics inputs, and cosmological
data~\GitHub. 

}

\author[a,b]{Quan-feng Wu \orcidlink{0000-0002-5716-5266}}
\author[a]{and Xun-Jie Xu \orcidlink{0000-0003-3181-1386}}
\affiliation[a]{Institute of High Energy Physics, Chinese Academy of Sciences, Beijing 100049, China}
\affiliation[b]{Kaiping Neutrino Research Center, Kaiping 529386, China}
\emailAdd{wuquanfeng@ihep.ac.cn} 
\emailAdd{xuxj@ihep.ac.cn} 
\preprint{\today}  
\maketitle

\section{Introduction}

Primordial black holes (PBHs) provide a unique window into the physics
of the early Universe. Unlike astrophysical black holes, which form
from the gravitational collapse of massive stars, PBHs could have
originated from the collapse of density fluctuations in the very early
Universe, well before the onset of structure formation~\cite{Escriva:2022duf}.
Their possible existence has been considered for decades \cite{Hawking:1971ei,Carr:1974nx,Carr:1975qj},
and in recent years they have attracted renewed interest due to the
rich phenomenology associated with them, including possible connections
to dark matter~\cite{Carr:2020xqk,Green:2020jor,Baldes:2020nuv,Gondolo:2020uqv,Bernal:2020kse,Bernal:2020bjf,Bernal:2020ili,Auffinger:2020afu,Cheek:2021odj,Masina:2021zpu,Cheek:2021cfe,Bernal:2021yyb,Bernal:2021bbv,Calabrese:2021src,Sandick:2021gew,Bernal:2022oha,Gehrman:2023esa,Arcadi:2024tib},
baryogenesis~\cite{Hooper:2020otu,Datta:2020bht,JyotiDas:2021shi,Bernal:2022pue,Coleppa:2022pnf,Gehrman:2022imk,Calabrese:2023key,Schmitz:2023pfy},
gravitational waves~\cite{Nakama:2016gzw,Garcia-Bellido:2016dkw,Raidal:2017mfl,Papanikolaou:2020qtd,Balaji:2024hpu,Domenech:2024wao},
and high-energy cosmic neutrinos~\cite{Wu:2024uxa,Zantedeschi:2024ram,Chianese:2024rsn,Klipfel:2025jql,Boccia:2025hpm,Choi:2025hqt,Anchordoqui:2025xug,Dvali:2025ktz,Dondarini:2025ktz}.

A key property of black holes is their instability under quantum effects.
As first demonstrated by Hawking~\cite{Hawking:1974rv}, black holes
radiate thermally and lose mass over time, a process that eventually
leads to complete evaporation on sufficiently long timescales.  The
lifetime of a black hole scales as the cube of its mass, so PBHs with
masses below about $10^{15}\,\mathrm{g}$ would have fully evaporated
by the present epoch. In particular, PBHs lighter than about $10^{9}\,\mathrm{g}$
would have completely evaporated before the onset of Big Bang nucleosynthesis
(BBN),  at cosmic times $t\lesssim1\,\mathrm{s}$.

BBN is among the most important probes of the early Universe, providing
precise predictions for the primordial abundances of light elements
such as deuterium, helium, and lithium. These predictions, derived
within the standard cosmological framework, are in remarkable agreement
with astronomical observations, leaving little room for significant
deviations in the thermal and particle history of the Universe around
the BBN epoch. As a result, any nonstandard physics that could inject
energy, alter particle abundances, or modify the expansion history
around  this epoch is subject to stringent constraints---see, e.g.,
Refs.~\cite{Kawasaki:2017bqm,Hufnagel:2017dgo,Hufnagel:2018bjp,Huang:2017egl,Depta:2019lbe,Sabti:2020yrt,Boyarsky:2020dzc,Chen:2024cla,Dev:2025pru}.

PBH evaporation prior to BBN is an especially interesting case in
this context. Although such PBHs would not survive long enough to
directly influence later cosmological processes, the particles and
entropy they release could still affect the conditions under which
nucleosynthesis begins.  Consequently, PBHs evaporating before BBN
have the potential to leave measurable imprints on the abundances
of light elements, thereby allowing BBN to serve as a sensitive probe
of their existence and abundance.

In this work, we investigate in detail how PBHs evaporating before
BBN could alter BBN predictions and derive the corresponding constraints
on such PBHs. Previous studies have presented several BBN constraints
on  PBHs~\cite{Carr:2020gox,Carr:2009jm,Keith:2020jww,Kohri:1999ex},
but the results in the literature exhibit significant discrepancies,
which necessitate a careful examination in a transparent framework.
Therefore, our work presents an anatomy of the calculation. We scrutinize
various steps that could affect the BBN sensitivity to PBHs, including
the background effect evaluation, the hadronization of Hawking radiation,
meson-driven neutron--proton conversion processes, and the evolution
of the neutron-to-proton ratio. At each step, we provide the output
of our code and carefully cross-check it against analytical estimates
and qualitative expectations.  This  allows us to disentangle the
physical effects responsible for modifying the light-element abundances
and to assess the robustness of the resulting BBN constraints. Finally,
we make our code publicly available, so that it can be readily improved
and extended to incorporate future updates in nuclear reaction rates,
particle physics inputs, and cosmological observations.

The structure of this paper is as follows. In Sec.~\ref{sec:formulation},
we introduce the basic formulations, including the standard BBN framework
and the properties of PBHs. In Sec.~\ref{sec:background}, we discuss
the influence of PBHs on the background evolution. Sec.~\ref{sec:hadronization}
 is devoted to the hadronization of Hawking radiation and meson-driven
neutron--proton conversion. In Sec.~\ref{sec:results}, we present
numerical solutions and results. Finally, we conclude in Sec.~\ref{sec:conclusion}
and relegate some details to the appendices.

\section{Basic formulations\label{sec:formulation}}

 We begin with a brief review of the relevant physics and formulae
involved in BBN and PBHs. 

\subsection{BBN}

When the temperature of the Universe, $T$, is above a few MeV, neutrons
and protons maintain chemical equilibrium with each other via weak
interaction processes:
\begin{align*}
n+\nu_{e} & \leftrightarrows p+e^{-}\thinspace,\\
n+e^{+} & \leftrightarrows p+\overline{\nu_{e}}\thinspace,\\
n & \leftrightarrows p+e^{-}+\overline{\nu_{e}}\thinspace.
\end{align*}
In the standard scenario, the chemical potentials of neutrinos ($\nu_{e}$,
$\overline{\nu_{e}}$) and electrons ($e^{\pm}$) are negligible.
Hence, the chemical equilibrium implies that the neutron and proton
chemical potentials, denoted by $\mu_{n}$ and $\mu_{p}$, are equal.
Since neutrons and protons are highly non-relativistic and sparse
at this temperature, they approximately obey the Maxwell-Boltzmann
statistics, with number densities given by 
\begin{equation}
n_{N}=e^{\mu_{N}/T}\frac{g_{N}m_{N}^{2}T}{2\pi^{2}}K_{2}\left(\frac{m_{N}}{T}\right),\label{eq:Nn}
\end{equation}
where $N\in\{n,\ p\}$, $g_{N}=2$ accounts for the  spin multiplicity,
$m_{N}$ denotes the mass of the nucleon $N$, and $K_{2}$ is the
modified Bessel function of the second kind.   Equation~\eqref{eq:Nn}
implies the following in-equilibrium ratio of the neutron and proton
number densities:
\begin{equation}
\frac{n_{n}}{n_{p}}\approx e^{-Q/T}\label{eq:}
\end{equation}
with $Q\equiv m_{n}-m_{p}\approx1.3$ MeV. 

At $T\ll Q$, the in-equilibrium ratio is exponentially suppressed.
However, the weak interaction processes can no longer maintain the
equilibrium when $T$ drops below about $1$ MeV,  and the neutrons
begin to freeze out.  The freeze-out process is quantitatively governed
by the  Boltzmann equation: 
\begin{equation}
\frac{dX_{n}}{dt}=-\Gamma_{n\to p}X_{n}+\Gamma_{p\to n}\left(1-X_{n}\right),\label{eq:Xn-ode}
\end{equation}
with
\begin{equation}
X_{n}\equiv\frac{n_{n}}{n_{n}+n_{p}}\thinspace,\label{eq:-1}
\end{equation}
where $\Gamma_{n\to p}$ and $\Gamma_{p\to n}$ denote the conversion
rates of a neutron to a proton and a proton to a neutron, respectively.
In the SM, $\Gamma_{n\to p}$ and $\Gamma_{p\to n}$ receive contributions
(denoted by $\Gamma_{n\to p}^{({\rm SM})}$ and $\Gamma_{p\to n}^{({\rm SM})}$)
from the aforementioned three weak processes. Under certain approximations
(neglecting quantum statistical effects and the electron mass), 
$\Gamma_{n\to p}^{({\rm SM})}$ and $\Gamma_{p\to n}^{({\rm SM})}$
can be calculated analytically, resulting in simple and compact expressions---see,
e.g., Ref.~\cite{Baumann:2022mni} or Appendix~\ref{sec:conversion-rates}.
 After including quantum statistical effects and the electron mass,
the calculation is more complicated and requires numerical integration---see,
e.g., Refs.~\cite{Meador-Woodruff:2024due,Kawano:1992ua}. In Appendix~\ref{sec:conversion-rates},
we compare the rates obtained via the analytical and numerical approaches
and demonstrate that the difference is negligibly small. 

In the presence of new physics, the conversion rates $\Gamma_{n\to p}$
and $\Gamma_{p\to n}$ may receive additional contributions. This
would directly affect the evolution of $X_{n}$ according to Eq.~\eqref{eq:Xn-ode}.
New physics could also play a more hidden role in Eq.~\eqref{eq:Xn-ode}
by modifying the cosmological expansion,  thereby altering the relation
between $T$ and $t$ (Note  that $\Gamma_{n\to p}$ and $\Gamma_{p\to n}$
are $T$-dependent). In  standard BBN which undergoes in the radiation-dominated
Universe, $t$ and $T$ are related by
\begin{align}
t & \approx\frac{1}{2H}\thinspace,\label{eq:-2}\\
H & \approx1.66g_{\star}^{1/2}\frac{T^{2}}{m_{{\rm pl}}}\thinspace,\label{eq:-3}
\end{align}
where $H$ is the Hubble parameter, $g_{\star}$ is the effective
number of degrees of freedom (decreasing from $10.75$ to $3.4$ when
$T$ decreases from 10 MeV to keV), and $m_{\text{pl}}\approx1.22\times10^{19}$
GeV is the Planck mass. If a non-radiation component  contributes
significantly to the total energy density of the Universe, both Eqs.~\eqref{eq:-2}
and \eqref{eq:-3} should be modified. If there is extra radiation,
Eq.~\eqref{eq:-3} should be modified. 

By solving Eq.~\eqref{eq:Xn-ode}, one obtains the evolution of $X_{n}$,
which is then used in the calculation of nucleosynthesis to determine
the relic abundances of light elements such as $\text{D}$, $^{3}\text{He}$,
$^{4}\text{He}$, and $^{7}\text{Li}$. 

It is important to note that below the MeV scale, the temperature
of neutrinos, $T_{\nu}$, starts to deviate from the temperature of
photons, $T_{\gamma}$, due to neutrino decoupling and $e^{+}e^{-}$
annihilation.  Throughout this work, we use $T$ generically when
this distinction is negligible; otherwise, $T$ should be understood
as $T_{\gamma}$. Using entropy conservation, $T_{\gamma}$ and the
scale factor $a$ are related by 
\begin{equation}
g_{\star,s}(T_{\gamma})a^{3}T_{\gamma}^{3}={\rm constant}\thinspace,\label{eq:-25}
\end{equation}
where  $g_{\star,s}$ is the effective number of degrees of freedom
in entropy. Appendix~\ref{sec:T-split} explains the numerical details
of how we determine $T_{\gamma}$, $T_{\nu}$, and $a$ from one to
another.

\subsection{PBHs}

A PBH, once formed in the early Universe, would keep evaporating via
Hawking radiation. This gradually reduces its mass, $m_{{\rm BH}}$,
 and increases its temperature, $T_{{\rm BH}}$, which is related
to its mass by
\begin{equation}
T_{{\rm BH}}=\frac{m_{{\rm pl}}^{2}}{8\pi m_{{\rm BH}}}\thinspace.\label{eq:-5}
\end{equation}
If the PBH  forms at $t=t_{i}$ with an initial mass $m_{{\rm BH},i}$,
the mass $m_{{\rm BH}}$  subsequently decreases as follows~\cite{Baldes:2020nuv}:
\begin{equation}
m_{{\rm BH}}=m_{{\rm BH},i}\left(1-\frac{t-t_{i}}{\tau_{{\rm BH}}}\right)^{\frac{1}{3}},\ \ t\in\left[t_{i},\ t_{{\rm ev}}\right],\label{eq:-6}
\end{equation}
where $\tau_{{\rm BH}}$ is the PBH lifetime, and $t_{{\rm ev}}$
is the evaporation time, determined by $t_{{\rm ev}}=t_{i}+\tau_{{\rm BH}}$.
The lifetime is given by~\cite{Baldes:2020nuv,Cheek:2021odj}
\begin{equation}
\tau_{{\rm BH}}=\frac{m_{{\rm BH},i}^{3}}{3g_{{\rm BH}}m_{{\rm pl}}^{4}}\thinspace,\ \ \text{with}\ \ g_{{\rm BH}}=\frac{27}{4}\times\frac{g_{\star}}{30720\pi}\thinspace,\label{eq:-7}
\end{equation}
where $g_{\star}$ should account for all particles lighter than $T_{{\rm BH}}$.
For PBHs considered in this work, $T_{{\rm BH}}$ is well above the
electroweak scale, allowing us to constantly set $g_{\star}=106.75$
in Eq.~\eqref{eq:-7}. For PBHs formed via collapse of local overdensities,
the formation time is 
\begin{equation}
t_{i}\approx\frac{m_{{\rm BH},i}}{\gamma m_{{\rm pl}}^{2}}\thinspace,\label{eq:-24}
\end{equation}
where $\gamma\approx0.2$ is a numerical factor quantifying the fraction
of mass within the particle horizon that gravitationally collapses
into the PBH~\cite{Carr:1975qj,Green:2004wb} (see also \cite{Carr:2009jm}
for more detailed discussions).   From this and Eq.~\eqref{eq:-7},
one can estimate the temperatures of the Universe at $t_{i}$ and
$t_{{\rm ev}}$, denoted by $T_{i}$ and $T_{{\rm ev}}$, respectively:
\begin{align}
T_{i} & =1.4\times10^{11}\text{GeV}\cdot\left(\frac{10^{9}\text{g}}{m_{{\rm BH},i}}\right)^{\frac{1}{2}},\label{eq:-8}\\
T_{{\rm ev}} & =1.8\ \text{MeV}\cdot\left(\frac{10.75}{g_{\star,\text{ev}}}\right)^{\frac{1}{4}}\cdot\left(\frac{10^{9}\text{g}}{m_{{\rm BH},i}}\right)^{\frac{3}{2}},\label{eq:-9}
\end{align}
where $g_{\star,\text{ev}}$ denotes $g_{\star}$ at evaporation. 

According to Eq.~\eqref{eq:-9}, PBHs with $m_{{\rm BH},i}\lesssim10^{9}$
gram evaporate completely  while all SM particles are still in equilibrium.
 Therefore, for PBHs to affect BBN significantly, the initial mass
needs to be higher than $10^{9}$ gram~\footnote{See, however, Ref.~\cite{Aljazaeri:2025ftv} for the potential impact
of early mergers which might alter this slightly.}. Above this  threshold,  PBHs can have the following effects on
BBN:

\begin{itemize}
\item \emph{Modifying the background:} PBHs and their emmited particles
 may constitute a significant amount of the energy of the Universe,
thereby  altering the Hubble expansion during the BBN epoch.
\item \emph{Modifying the ingredients:} Through certain reaction processes,
particles emitted by PBHs may affect the abundance of neutrons ($n$)
and protons ($p$) before these BBN ingredients are fused into  light
elements ($\text{D}$, $^{3}\text{He}$, $^{4}\text{He}$, $^{7}\text{Li}$).
\item \emph{Modifying the products:} Energetic radiation from PBHs can dissociate
the light elements produced by nucleosynthesis, typically reducing
the $^{4}\text{He}$ abundance and increasing the $\text{D}$ abundance. 
\end{itemize}

The first effect is subdominant compared to the second and will be
estimated  in the next section. The second effect turns out to be
the most important for PBHs  that evaporate before BBN and is therefore
the primary focus of this work. The third effect becomes relevant
only for PBHs  that evaporate after BBN, where a proper treatment
requires  dedicated calculations of photo-dissociation and hadro-dissociation
processes. Such an analysis is beyond the scope of this work and 
left  to future work.

\section{Influence on the background\label{sec:background}}

In this section, we estimate the influence of PBH evaporation on the
background evolution of the Universe. The abundance of PBHs is usually
parametrized by 
\begin{equation}
\beta\equiv\frac{\rho_{{\rm BH}}(t_{i})}{\rho_{{\rm tot}}(t_{i})}\thinspace,\label{eq:-20}
\end{equation}
where $\rho_{{\rm BH}}$ and $\rho_{{\rm tot}}$ denote the energy
densities of the PBHs and the Universe, respectively. Using $\rho_{{\rm tot}}=3m_{{\rm pl}}^{2}H^{2}/(8\pi)$
and Eq.~\eqref{eq:-24} with $t=1/(2H)$, one  obtains
\begin{equation}
n_{{\rm BH},i}=\frac{3\beta\gamma^{2}m_{{\rm pl}}^{6}}{32\pi m_{{\rm BH},i}^{3}}\thinspace.\label{eq:-26}
\end{equation}
Since the total number of PBHs in a comoving volume is conserved 
for $t_{i}<t<t_{{\rm ev}}$, the number density evolves as
\begin{equation}
n_{{\rm BH}}=n_{{\rm BH},i}\left(\frac{a_{i}}{a}\right)^{3}=n_{{\rm BH},i}\left(\frac{T_{\gamma}}{T_{i}}\right)^{3}\frac{g_{\star,s}(T_{\gamma})}{g_{\star,s}(T_{i})}\thinspace.\label{eq:-29}
\end{equation}
Hence,
\begin{equation}
\rho_{\text{BH}}\approx n_{{\rm BH}}m_{{\rm BH},i}\approx\frac{3\beta\gamma^{2}m_{{\rm pl}}^{6}}{32\pi m_{{\rm BH},i}^{2}}\cdot\left(\frac{T_{\gamma}}{T_{i}}\right)^{3}\frac{g_{\star,s}(T_{\gamma})}{g_{\star,s}(T_{i})}\ \ ({\rm for}\ t\ll t_{{\rm ev}})\thinspace.\label{eq:-30}
\end{equation}
Let us compare it to the energy density of neutrinos of a single flavor,
\begin{equation}
\rho_{\nu}=\frac{\pi^{2}}{30}\times2\times\frac{7}{8}T_{\nu}^{4}\thinspace,\label{eq:-27}
\end{equation}
which gives
\begin{equation}
\frac{\rho_{\text{BH}}}{\rho_{\nu}}\approx0.040\cdot\frac{\beta}{10^{-16}}\cdot\left(\frac{10^{9}\ {\rm g}}{m_{{\rm BH},i}}\right)^{\frac{1}{2}}\cdot\left(\frac{2\text{MeV}}{T}\right)\cdot\left(\frac{T_{\gamma}}{T_{\nu}}\right)^{4}\cdot\frac{g_{\star,s}(T)}{10.75}\thinspace.\label{eq:-28}
\end{equation}
This ratio grows as $T$ decreases (i.e. as the Universe expands)
until $T$ reaches $T_{{\rm ev}}$ given in Eq.~\eqref{eq:-9}. The
maximum of the ratio can be roughly estimated by substituting $T=T_{{\rm ev}}$
into Eq.~\eqref{eq:-28}, yielding
\begin{equation}
\left(\frac{\rho_{\text{BH}}}{\rho_{\nu}}\right)_{{\rm max}}\approx0.044\cdot\frac{\beta}{10^{-16}}\cdot\left(\frac{m_{{\rm BH},i}}{10^{9}\ {\rm g}}\right),\label{eq:-31}
\end{equation}
where we have assumed that  PBHs evaporate before $e^{+}e^{-}$ annihilation,
allowing us to set $T_{\gamma}=T_{\nu}$ and $g_{\star,s}(T_{{\rm ev}})=g_{\star}(T_{{\rm ev}})=10.75$.
If they evaporate after that, these relations are modified but the
resulting coefficient only changes slightly from $0.044$ to $0.046$,
 a negligible difference for our discussions. 

\begin{figure}
\centering

\includegraphics[width=0.8\textwidth]{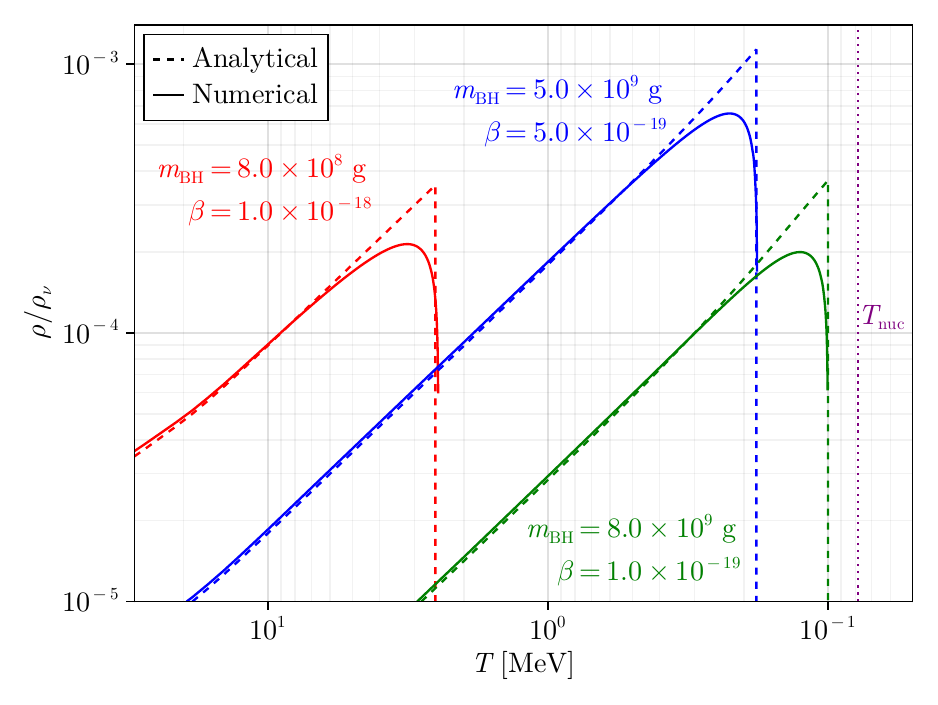}

\caption{The energy density of PBHs $\rho_{\text{BH}}$ compared with the neutrino
energy density $\rho_{\nu}$. Here solid lines are obtained by numerically
solving the evolution of $\rho_{\text{BH}}$; dashed lines represent
the analytical estimate in Eq.~\eqref{eq:-28}. The vertical dotted
line indicates the temperature of nucleosynthesis.  \label{fig:rho}
}

\end{figure}

Figure~\ref{fig:rho} shows the numerical evolution of $\rho_{\text{BH}}/\rho_{\nu}$,
which is close to the analytical estimate  in Eq.~\eqref{eq:-28}.
The vertical cut-off is estimated using Eq.~\eqref{eq:-9} and the
vertical dotted line indicates the temperature of nucleosynthesis,
$T_{{\rm nuc}}\approx0.07$ MeV. 

When PBHs evaporate after neutrino decoupling, they inject energy
and entropy into two sectors: neutrinos and the electromagnetically
coupled thermal plasma which eventually reduces to photons. The most
restrictive constraint on such an effect comes from the effective
number of neutrino species, $N_{{\rm eff}}$, which would increase
or decrease if the energy is mainly injected into neutrinos or photons,
respectively. $N_{{\rm eff}}$ has been measured very precisely by
 both CMB and BBN observations. The latest CMB  measurement is $N_{{\rm eff}}=2.99\pm0.17$~\cite{Planck:2018vyg},
and a recent BBN analysis gives $N_{{\rm eff}}=2.898\pm0.141$~\cite{Yeh:2022heq}.
The measurements have been used to set stringent constraints on relevant
new physics~\cite{Escudero:2018mvt,Escudero:2019gzq,Abazajian:2019oqj,EscuderoAbenza:2020cmq,Luo:2020sho,Luo:2020fdt,Yeh:2022heq,Li:2022dkc,Li:2023puz,Ovchynnikov:2024xyd,Ovchynnikov:2024rfu}. 

The fraction of PBH energy transferred to neutrinos is complicated
to evaluate, due to secondary productions from particle decays. It
is, however, straightforward to estimate an upper bound on the impact
by assuming that all the energy goes into neutrinos. In this way,
Eq.~\eqref{eq:-31} can be interpreted as the maximal contribution
to  $N_{{\rm eff}}$. Consequently, $N_{{\rm eff}}$ is  increased
at most by
\begin{equation}
\Delta N_{{\rm eff}}\lesssim0.044\cdot\frac{\beta}{10^{-16}}\cdot\left(\frac{m_{{\rm BH},i}}{10^{9}\ {\rm g}}\right).\label{eq:-32}
\end{equation}
Since $0.044$ is beyond the current sensitivity  of BBN and CMB
observations, PBHs with $\beta\lesssim10^{-16}\times\left(10^{9}\ {\rm g}/m_{{\rm BH},i}\right)$
cannot influence the cosmological background significantly. In the
next section, we will show that products of PBH evaporation directly
participating in hadronic interactions with BBN ingredients have a
much stronger impact on BBN. We note here that the above estimate
only applies to PBHs that evaporate after neutrino decoupling, corresponding
to $m_{{\rm BH},i}\gtrsim10^{9}$ g. For smaller PBHs evaporating
before neutrino decoupling, almost all effects caused by such PBHs
are washed out by thermal interactions, except for the dilution effect
on $\eta_{B}$~\cite{Boccia:2024nly}. However, since $\eta_{B}$
is determined from observations at later times rather than fixed {\it
a priori}, this effect should not be used to impose a valid constraint.

\section{Influence on the BBN ingredients via hadronic interactions \label{sec:hadronization}}

\subsection{Hadronization of Hawking radiation}

\begin{figure}
\centering

\includegraphics[width=0.7\textwidth]{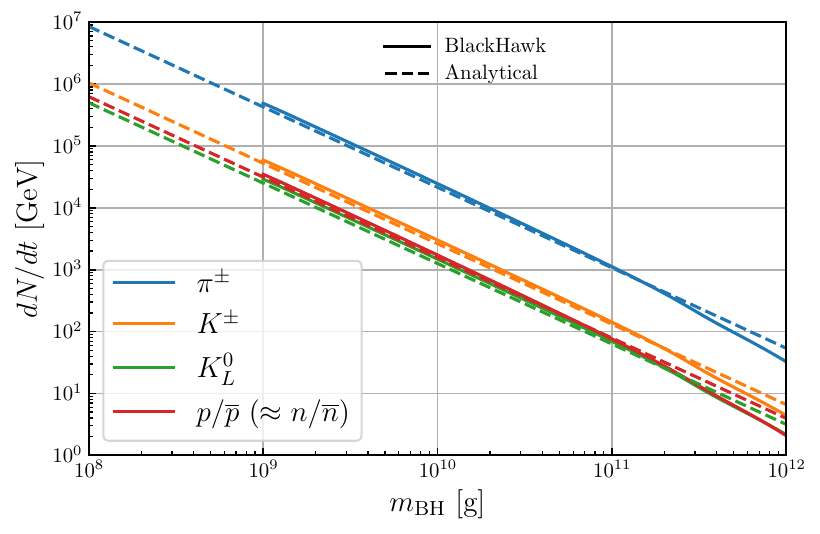}

\caption{\label{fig:hadron}Hadronic  production rates of PBHs obtained from
{\tt BlackHawk} (solid lines) and analytical estimates (dashed lines).
The solid lines are not extended to $m_{{\rm BH}}<10^{9}$ g because
below this mass, we are unable to obtain {\tt BlackHawk} results
unaffected by the energy limit imposed on partons---see the main
text for details.  }
\end{figure}

Through Hawking radiation, PBHs emit various elementary particles
in the SM, including quarks, leptons, gauge bosons, etc. Many of these
particles decay rapidly after production, ending up with stable particles
such as photons, neutrinos, and electrons. Light quarks and gluons
hadronize, producing a large amount of hadrons like pions ($\pi^{\pm}$,
$\pi^{0}$), kaons ($K^{\pm}$, $K_{L,S}^{0}$), and nucleons/anti-nucleons
($n$/$\overline{n}$, $p$/$\overline{p}$). The mesons produced
by PBHs are responsible for meson-driven $n\leftrightarrow p$ conversion.
The anti-nucleons can annihilate with nucleons, tending to drive $X_{n}$
toward $1/2$.  

In principle, neutrinos and electrons produced by PBHs could affect
$n\leftrightarrow p$ conversion processes. However, the cross sections
of $\nu_{e}+n\leftrightarrow p+e^{-}$ and $e^{+}+n\leftrightarrow p+\overline{\nu_{e}}$,
typically around $10^{-43}\ \text{cm}^{2}$, are orders of magnitude
lower than those of meson-driven processes, which are at least above
one millibarn (mb, $1\ \text{mb}\equiv10^{-27}\ \text{cm}^{2}$).
Given that the emissivity of leptons from a high-temperature PBH is
lower than the emissivity of mesons, the latter plays a significantly
more important role than the former in BBN. 

To quantitatively assess the impact  of the hadrons, we need to
handle the hadronization of Hawking radiation  properly. Although
this is implemented in the package \texttt{BlackHawk}~\cite{Arbey:2019mbc,Arbey:2021mbl},
we shall bring a minor issue to the reader's attention.  \texttt{BlackHawk}
(v2.3) applies the hadronization tables generated by \texttt{C++}
code that invokes \texttt{PYTHIA} (v8)~\cite{Sjostrand:2019zhc,Bierlich:2022pfr}.
The \texttt{C++} code contains ``\texttt{const double Emax\_init = 100000.;}'',
restricting the maximum parton energy in hard processes to $10^{5}$
GeV. Hence {\tt BlackHawk} by default should not be used to compute
very energetic meson fluxes above this energy scale. In practice,
we find that PBHs with $m_{{\rm BH},i}\gtrsim10^{9}$ g are not significantly
affected by this limitation. One could modify {\tt BlackHawk} to
 extend it to higher energies manually. However, since {\tt PYTHIA}
is a Monte-Carlo event generator, the running time increases drastically
at energies higher than that. 

In this work, we obtain the meson fluxes by running {\tt BlackHawk}
(solid lines in Fig.~\ref{fig:hadron}) combined with analytical
estimates (dashed lines in Fig.~\ref{fig:hadron}). More specifically,
the emissivity of a secondary species such as mesons can be estimated
by~\cite{MacGibbon:1990zk,MacGibbon:1991tj}
\begin{equation}
\frac{d^{2}N_{j}}{dtdE_{j}}=\sum_{i}\frac{d^{2}N_{\text{BH}\to i}}{dtdE_{i}}\frac{dN_{i\to j}}{dE_{j}}dE_{i}\thinspace,\label{eq:-33}
\end{equation}
where $i$ and $j$ denote the primary and secondary particle species,
$N_{\text{BH}\to i}$ and $N_{i\to j}$ represent the number of $i$
emitted by a PBH and the number of $j$ generated by an $i$ particle,
respectively. Since hadrons produced by PBHs in the early Universe
lose energy rapidly in the dense $\gamma$-$e^{\pm}$ plasma\footnote{During the epoch from neutrino decoupling to BBN, all particles with
energies much higher than $T$ can lose energy efficiently via elastic
scattering or Cherenkov radiation and reach kinetic equilibrium rapidly,
except for neutrinos. Neutrinos emitted by PBHs in the early Universe
may contribute to the high-energy and ultra-high-energy neutrino fluxes
 observed today by neutrino telescopes~\cite{Wu:2024uxa}.}, we are mainly concerned with the total numbers of the emitted mesons
instead of their energy distributions. Hence we integrate out $E_{j}$
in Eq.~\eqref{eq:-33} and focus on 
\begin{equation}
\frac{dN_{j}}{dt}=\sum_{i}\frac{d^{2}N_{\text{BH}\to i}}{dtdE_{i}}N_{i\to j}dE_{i}\thinspace.\label{eq:-34}
\end{equation}

The Hawking radiation rate, $\frac{d^{2}N_{\text{BH}\to i}}{dtdE_{i}}$,
is calculated as follows~\cite{Page:1976df}:
\begin{equation}
\frac{d^{2}N_{\text{BH}\to i}}{dtdE_{i}}=\frac{g_{i}}{2\pi}\frac{\gamma_{{\rm gray}}}{\exp\left(E_{i}/T_{{\rm BH}}\right)\pm1}\thinspace,\label{eq:-4}
\end{equation}
where $g_{i}$ denotes the multiplicity of particle $i$ being emitted
and $\gamma_{{\rm gray}}$ is the graybody factor. In the geometric
optics limit, $\gamma_{{\rm gray}}\approx27E_{i}^{2}m_{{\rm BH}}^{2}/m_{{\rm pl}}^{4}$.
Deviations of $\gamma_{{\rm gray}}$ from the geometric optics limit
can be acquired from Fig.~1 in Ref.~\cite{Cheek:2021odj} or Fig.~1
of Ref.~\cite{Arbey:2021mbl}. 

For hadrons produced by the hadronization of gluons ($i=g$) and light
quarks ($i=q$), $N_{i\to j}$ depends on the parton energy $E_{i}$.
The energy dependence in the limit of asymptotic freedom can be approximated
by\footnote{The same power law has also been noticed in previous studies~\cite{Keith:2020jww,Kawasaki:2017bqm}.}
\begin{equation}
N_{i\to j}(E_{i})\approx N_{i\to j}^{\star}\cdot\left(E_{i}/\text{TeV}\right)^{0.3},\label{eq:-35}
\end{equation}
where $N_{i\to j}^{\star}\equiv N_{i\to j}(1\ \text{TeV})$. We obtain
Eq.~\eqref{eq:-35} by running {\tt PYTHIA} (v8) independently of
{\tt BlackHawk}. The prefactor $N_{i\to j}^{\star}$ for a given
process is also determined with {\tt PYTHIA} (see Appendix \ref{sec:Pythia}
for details), from which  we obtain the following values:  $N_{q,g\to\pi^{\pm}}^{\star}=(54.7,\ 97.0)$,
$N_{q,g\to K^{\pm}}^{\star}=(6.5,\ 11.7)$, $N_{q,g\to K_{L}^{0}}^{\star}=(3.1,\ 5.7)$,
$N_{q,g\to p}^{\star}=(3.9,\ 6.8)$, and $N_{q,g\to n}^{\star}=(3.7,\ 7.0)$.
Note that these numbers include contributions of both particles and
antiparticles. 

Substituting Eqs.~\eqref{eq:-4} and \eqref{eq:-35} into Eq.~\eqref{eq:-34},
we arrive at
\begin{equation}
\frac{dN_{j}}{dt}=\left(9.24\ \text{GeV}\cdot g_{q}N_{q\to j}^{\star}+3.12\ \text{GeV}\cdot g_{g}N_{g\to j}^{\star}\right)\left(\frac{T_{{\rm BH}}}{\text{TeV}}\right)^{1.3}.\label{eq:-36}
\end{equation}
Here $g_{q}=3\times3\times2\times2=36$ and $g_{g}=8$ account for
the multiplicity of quarks and gluons, respectively. For quarks, it
includes colors, flavors, isospins, and spin polarizations. For gluons,
it includes colors. Note that in {\tt PYTHIA}, $q$ and $g$ are
always generated in pairs with opposite colors. So the factor of two
related to the pair production is canceled out by the factor of two
arising from quarks/antiquarks or gluon polarizations. 

Taking the specific values of $g_{q}$, $g_{g}$, and $N^{\star}$,
we plot Eq.~\eqref{eq:-36} in Fig.~\ref{fig:hadron} as dashed
lines, which approximately agree with the results obtained from {\tt
BlackHawk} (solid lines). Note that in Fig.~\ref{fig:hadron}, we
only present {\tt BlackHawk} results for $m_{{\rm BH}}\geq10^{9}$
g due to the $10^{5}$ GeV issue mentioned earlier. In our analysis,
when $m_{{\rm BH}}<10^{9}$ g, we use the analytical results and calibrate
them (by rescaling) to match the {\tt BlackHawk} results at $m_{{\rm BH}}=10^{9}$
g; when $m_{{\rm BH}}>10^{9}$ g, we use the {\tt BlackHawk} results.

It should be noted that while these mesons are continuously produced
by PBHs, they are concurrently depleted through decay or scattering
with nucleons. Since the lifetimes of these mesons ($\tau_{\pi^{\pm}}=2.6\times10^{-8}$
s, $\tau_{K^{\pm}}=1.2\times10^{-8}$ s, $\tau_{K_{L}^{0}}=5.1\times10^{-8}$
s) are much shorter than the relevant BBN time scales ($1\sim100$
s), their abundances would decline rapidly via decay if the production
ceased.  Owing to the continuous production, their abundances are
maintained at steady values determined by the following balance:
\begin{equation}
n_{j}\left(\Gamma_{j}^{\text{dec}}+\Gamma_{j}^{{\rm scat}}\right)\approx n_{{\rm BH}}\frac{dN_{j}}{dt}\thinspace,\label{eq:-37}
\end{equation}
where $j\in\{\pi^{\pm},\ K^{\pm},\ K_{L}^{0}\}$ and $\Gamma_{j}^{\text{dec/scat}}$
denotes the depletion rate of species $j$ through decay/scattering.
Equation~\eqref{eq:-37} implies that very short-lived mesons have
highly suppressed abundances. When $\Gamma_{j}^{\text{dec}}\gg\Gamma_{j}^{{\rm scat}}$,
$n_{j}$ determined from Eq.~\eqref{eq:-37} is proportional to $1/\Gamma_{j}^{\text{dec}}$,
i.e., to the lifetime of $j$. For this reason, very short-lived mesons
such as $\pi^{0}$ and $K_{S}^{0}$ are not included in our analysis.
For anti-protons and anti-neutrons ($j=\overline{p}$, $\overline{n}$),
Eq.~\eqref{eq:-37} also applies, with $\Gamma_{j}^{\text{dec}}=0$
and $\Gamma_{j}^{{\rm scat}}=n_{p}\langle\sigma v\rangle_{jp}+n_{n}\langle\sigma v\rangle_{jn}$,
where $\langle\sigma v\rangle_{jp}$ and $\langle\sigma v\rangle_{jn}$
are the baryon annihilation cross sections of $j$ with $p$ and $n$,
respectively. 

Equation~\eqref{eq:-37} is valid  when both $\Gamma_{j}\equiv\Gamma_{j}^{\text{dec}}+\Gamma_{j}^{{\rm scat}}$
and $\frac{dN_{j}}{dt}$ are well above the Hubble expansion rate
$H$. Hence, at high temperatures when $H\propto T^{2}$ exceeds either
$\Gamma_{j}$ or $\frac{dN_{j}}{dt}$, the actual evolution of the
hadron number densities must be determined by numerically solving
the relevant Boltzmann equations, which we have implemented in our
code.  Figure~\ref{fig:rho} shows the numerical solutions (solid
lines) alongside the analytical estimates (dashed lines) from Eq.~\eqref{eq:-37}.
 One can see that the hadron number densities computed from Eq.~\eqref{eq:-37}
are in excellent agreement with the numerical solutions. The meson
curves exhibit a plateau phase, which occurs when $\Gamma_{j}^{\text{dec}}\gg\Gamma_{j}^{{\rm scat}}$
and $dN_{j}/dt$ is approximately constant. We have checked that indeed
this plateau can be estimated by neglecting $\Gamma_{j}^{{\rm scat}}$
in Eq.~\eqref{eq:-37}, i.e., $n_{j}\approx n_{{\rm BH}}\frac{dN_{j}}{dt}/\Gamma_{j}^{\text{dec}}$. 

\begin{figure}
\centering

\includegraphics[width=0.7\textwidth]{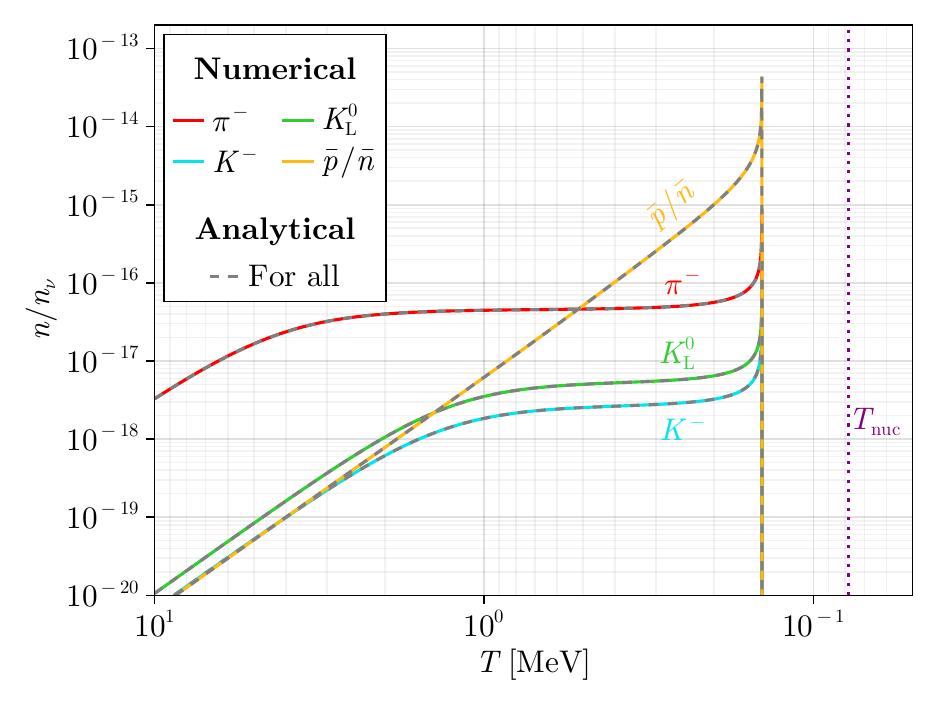}

\caption{\label{fig:hadron-evol} The number densities of hadrons obtained
by numerically solving the corresponding Boltzmann equations (solid
lines) compared with the analytical estimates using Eq.~\eqref{eq:-37}
(dashed lines).  The shown example assumes $m_{{\rm BH},i}=6\times10^{9}$
g and $\beta=10^{-16}$. }
\end{figure}

\subsection{Meson-driven $n\leftrightarrow p$ conversion}

Having determined the abundances of mesons emitted by PBHs, we can
now evaluate their impact on BBN. Mesons can  interact directly
with neutrons and protons, inducing  meson-driven $n\leftrightarrow p$
conversion processes. For instance, charged pions can convert $n$
and $p$ from one to the other through the following reactions:
\begin{align*}
\pi^{+}+n & \to p+\pi^{0}\thinspace,\\
\pi^{-}+p & \to n+\pi^{0}/\gamma\thinspace.
\end{align*}
Kaons can also induce similar conversions, but due to $m_{K^{0}}>m_{K^{\pm}}$
(unlike pions, for which the neutral one is lighter) with a relatively
large mass difference,  processes  such as $K^{-}+p\to n+K^{0}$
with $K^{0}$ in the final state make negligible contributions in
a MeV thermal bath. Instead, one should  include kaon reactions involving
$\Sigma^{\pm,0}$ and $\Lambda$ as intermediate states---see Ref.~\cite{Pospelov:2010cw}
for the details. 

These meson-driven $n\leftrightarrow p$ conversion processes can
be taken into account by adding their contributions to $\Gamma_{n\to p}$
and $\Gamma_{p\to n}$ in Eq.~\eqref{eq:Xn-ode}:
\begin{align}
\Gamma_{n\to p} & =\Gamma_{n\to p}^{({\rm SM})}+\sum_{j}\Gamma_{n\to p}^{j}\thinspace,\ \ \ \Gamma_{n\to p}^{j}\equiv n_{j}\langle\sigma_{n\to p}^{j}v\rangle\thinspace;\label{eq:-43}\\
\Gamma_{p\to n} & =\Gamma_{p\to n}^{({\rm SM})}+\sum_{j}\Gamma_{p\to n}^{j}\thinspace,\ \ \ \Gamma_{p\to n}^{j}\equiv n_{j}\langle\sigma_{p\to n}^{j}v\rangle\thinspace.\label{eq:-44}
\end{align}
Here $\langle\sigma_{n\to p}^{j}v\rangle$ and $\langle\sigma_{p\to n}^{j}v\rangle$
denote the thermally-averaged cross sections for the $n\to p$ and
$p\to n$ processes driven by meson $j$. Their specific values are
listed as follows~\cite{Kohri:2001jx}:
\begin{align}
\left\langle \sigma_{n\to p}^{\pi^{+}}v\right\rangle  & \approx1.7\ \text{mb}\thinspace,\label{eq:-38}\\
\left\langle \sigma_{p\to n}^{\pi^{-}}v\right\rangle /C_{\pi} & \approx1.5\ \text{mb}\thinspace,\label{eq:-39}\\
\left\langle \sigma_{n\to p}^{K^{-}}v\right\rangle  & \approx26\ \text{mb}\thinspace,\label{eq:-40}\\
\left\langle \sigma_{p\to n}^{K^{-}}v\right\rangle /C_{K} & \approx31\ \text{mb}\thinspace.\label{eq:-41}
\end{align}
Here $C_{\pi}$ and $C_{K}$ are Sommerfeld enhancement factors 
arising from the Coulomb attraction between oppositely charged particles.
These enhancement factors are computed by 
\begin{equation}
C_{j}=\frac{\epsilon_{j}}{1-e^{-\epsilon_{j}}}\thinspace,\ \ \text{with}\ \ \epsilon_{j}\equiv2\pi\alpha\sqrt{\frac{m_{j}m_{p}}{2T(m_{j}+m_{p})}}\thinspace,\label{eq:-42}
\end{equation}
 where $\alpha\approx1/137$ is the fine-structure constant. 

\subsection{Nucleon annihilation}

The hadronization of Hawking radiation generates the same amount of
nucleons ($p$, $n$) and antinucleons ($\overline{p}$, $\overline{n}$),
both reaching kinetic equilibrium with the thermal bath rapidly. If
$\overline{p}$ annihilates with $p$ in the thermal bath, it implies
one proton is removed from the thermal bath but meanwhile another
proton (originating from PBHs) is added to the thermal bath. So the
net effect of $\overline{p}p$ annihilation is equivalent to the pair
of PBH-generated $\overline{p}p$ ``decaying'', which only injects
a negligibly small energy into the thermal bath. If $\overline{p}$
annihilates with $n$ instead of $p$, then one neutron is removed
while another proton is added. In this case, the net effect is equivalent
to $n\to p$ conversion. Therefore, among the two possible annihilation
processes, only $\overline{p}n$ annihilation contributes to the conversion
rate while $\overline{p}p$ annihilation only plays the role of consuming
$\overline{p}$. For $\overline{n}$, there are also two similar annihilation
processes playing similar roles. 

The annihilation cross sections of these processes are given by~\cite{Kohri:1999ex}
\begin{align}
\langle\sigma v\rangle_{\overline{p}p} & \approx\langle\sigma v\rangle_{\overline{n}n}\approx37\ \text{mb}\thinspace,\label{eq:-62}\\
\langle\sigma v\rangle_{\overline{n}p} & \approx\langle\sigma v\rangle_{\overline{p}n}\approx28\ \text{mb}\thinspace.\label{eq:-63}
\end{align}
To include the effect of $\overline{p}n$ and $\overline{n}p$ annihilation
on $n\leftrightarrow p$ conversion, we add the following rates to
$\Gamma_{n\to p}$ and $\Gamma_{p\to n}$ in Eqs.~\eqref{eq:-43}
and \eqref{eq:-44}:
\begin{equation}
\Gamma_{n\to p}^{(\text{anni})}=n_{\overline{p}}\langle\sigma v\rangle_{\overline{p}n}\thinspace,\ \ \Gamma_{p\to n}^{(\text{anni})}=n_{\overline{n}}\langle\sigma v\rangle_{\overline{n}p}\thinspace.\label{eq:-64}
\end{equation}
Since the depletion rates of $\overline{p}$ and $\overline{n}$ 
are given by $n_{p}\langle\sigma v\rangle_{\overline{p}p}+n_{n}\langle\sigma v\rangle_{\overline{p}n}$
and $n_{n}\langle\sigma v\rangle_{\overline{n}n}+n_{p}\langle\sigma v\rangle_{\overline{n}p}$,
using Eq.~\eqref{eq:-37}, we obtain
\begin{align}
\Gamma_{n\to p}^{(\text{anni})} & \approx\frac{\langle\sigma v\rangle_{\overline{p}n}}{n_{p}\langle\sigma v\rangle_{\overline{p}p}+n_{n}\langle\sigma v\rangle_{\overline{p}n}}n_{{\rm BH}}\frac{dN_{j}}{dt}\thinspace,\label{eq:-65}\\
\Gamma_{p\to n}^{(\text{anni})} & \approx\frac{\langle\sigma v\rangle_{\overline{n}p}}{n_{n}\langle\sigma v\rangle_{\overline{n}n}+n_{p}\langle\sigma v\rangle_{\overline{n}p}}n_{{\rm BH}}\frac{dN_{j}}{dt}\thinspace,\label{eq:-66}
\end{align}
which implies that the two conversion rates would counteract each
other when $n_{n}\approx n_{p}$. That is, these two conversion rates
 tend to drive $n_{n}/n_{p}$ toward $1$, in contrast to the SM
conversion rates which drive $n_{n}/n_{p}$ toward $e^{-Q/T}$.

\section{Numerical solutions and results\label{sec:results}}

Solving Eq.~\eqref{eq:Xn-ode} with $\Gamma_{n\to p}$ and $\Gamma_{p\to n}$
including contributions from Eqs.~\eqref{eq:-43}, \eqref{eq:-44}
and \eqref{eq:-64}, we obtain the evolution of $X_{n}$ influenced
by PBH evaporation. Our code for numerically solving all relevant
differential equations is publicly available via GitHub\footnote{\GitHub*}.

\begin{figure}
\centering

\includegraphics[width=1\textwidth]{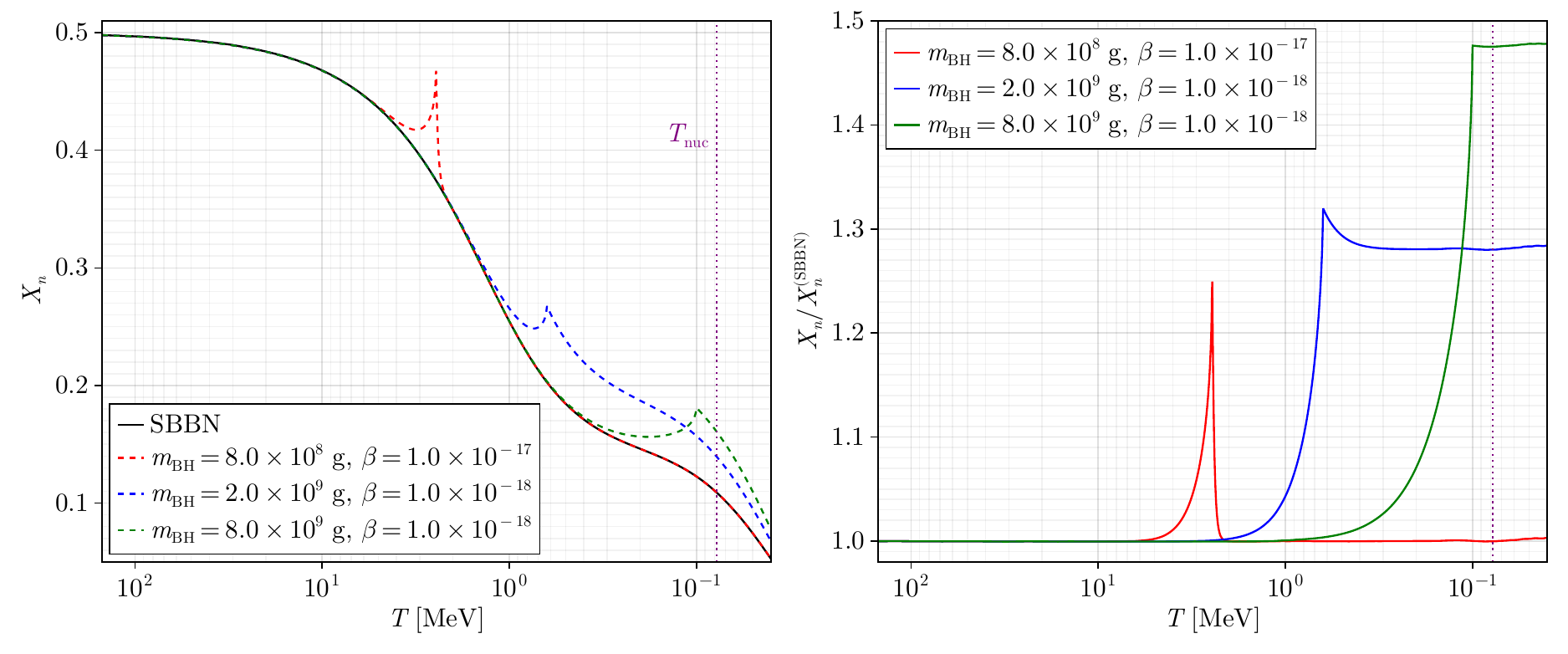}

\caption{Evolution of $X_{n}$ in the presence of PBH evaporation. The left
panel presents values of $X_{n}$ and the right panel shows the ratio
$X_{n}/X_{n}^{{\rm (SBBN)}}$ where $X_{n}^{({\rm SBBN)}}$ denotes
$X_{n}$ in standard BBN. \label{fig:sol}}
\end{figure}

In Fig.~\ref{fig:sol}, we demonstrate solutions for three benchmarks
with $\left(m_{{\rm BH}}/10^{9}\text{g},\ \beta\right)=(0.8,\ 10^{-17})$,
$(2,\ 10^{-17})$, and $(8,\ 10^{-16})$. The left panel shows the
evolution of $X_{n}$, while the right panel shows the ratio $X_{n}/X_{n}^{{\rm (SBBN)}}$
where $X_{n}^{({\rm SBBN)}}$ denotes $X_{n}$ in the standard BBN
framework. 

From the left panel of Fig.~\ref{fig:sol}, one can see that when
PBHs of a certain mass evaporates, they produce a spike on the $X_{n}$
curve. This occurs because the burst of mesons or antinucleons emitted
by  PBHs near the end of their lifetimes convert a considerably
large amount of protons to neutrons. The location of the spike can
be estimated using Eq.~\eqref{eq:-9} and the height of the spike
is sensitive to $\beta$.    

The spike enhances $X_{n}$ significantly when the evaporation completes.
Depending on whether this happens before or after neutron freeze-out
(which happens roughly at $T\sim1$ MeV), the impact may be washed
out by the thermal interactions of nucleons ($n$, $p$) with leptons
($\nu$, $e$). For instance, the red curve  shows a spike occurring
above 2 MeV, so its impact is rapidly washed out, leaving the subsequent
evolution of $X_{n}$ nearly identical to the standard scenario.
The blue curve shows a spike at about 0.6 MeV, where the wash-out
effect is still present but much weaker. So after the spike, there
is a modest reduction in the ratio $X_{n}/X_{n}^{{\rm (SBBN)}}$
caused by the wash-out effect, as shown in the right panel of Fig.~\ref{fig:sol}.
After a short period of washing out, $X_{n}/X_{n}^{{\rm (SBBN)}}$
reaches a steady value. The green curve is generated by much slower
PBH evaporation, rendering the spike less significant. The evaporation
completes at $T=0.1$ MeV, a temperature sufficiently low to suppress
the wash-out effect. So its $X_{n}/X_{n}^{{\rm (SBBN)}}$ curve barely
decreases after the spike. 

 The quantity most relevant to BBN observables is the value of $X_{n}$
immediately before nucleosynthesis, $X_{n}(T_{{\rm nuc}})$,  
from which the $^{4}\text{He}$ mass fraction $Y_{P}$ can be directly
determined by
\begin{equation}
Y_{P}\approx2X_{n}(T_{{\rm nuc}})\thinspace.\label{eq:-60}
\end{equation}
Equation~\eqref{eq:-60}  holds approximately because most of the
neutrons present at $T=T_{{\rm nuc}}$ are ultimately incorporated
into $^{4}\text{He}$. This allows us to determine $Y_{P}$ without
solving the full set of coupled differential equations in the nuclear
reaction network, which is  more computationally expensive and more
susceptible to numerical instabilities. In our work, we calculate
$Y_{P}$ using both approaches (the nuclear reaction network is implemented
by importing several  modules from {\tt BBN-simple}~\cite{Meador-Woodruff:2024due})
and find close agreement between the corresponding results.

\begin{figure}
\centering

\includegraphics[width=0.7\textwidth]{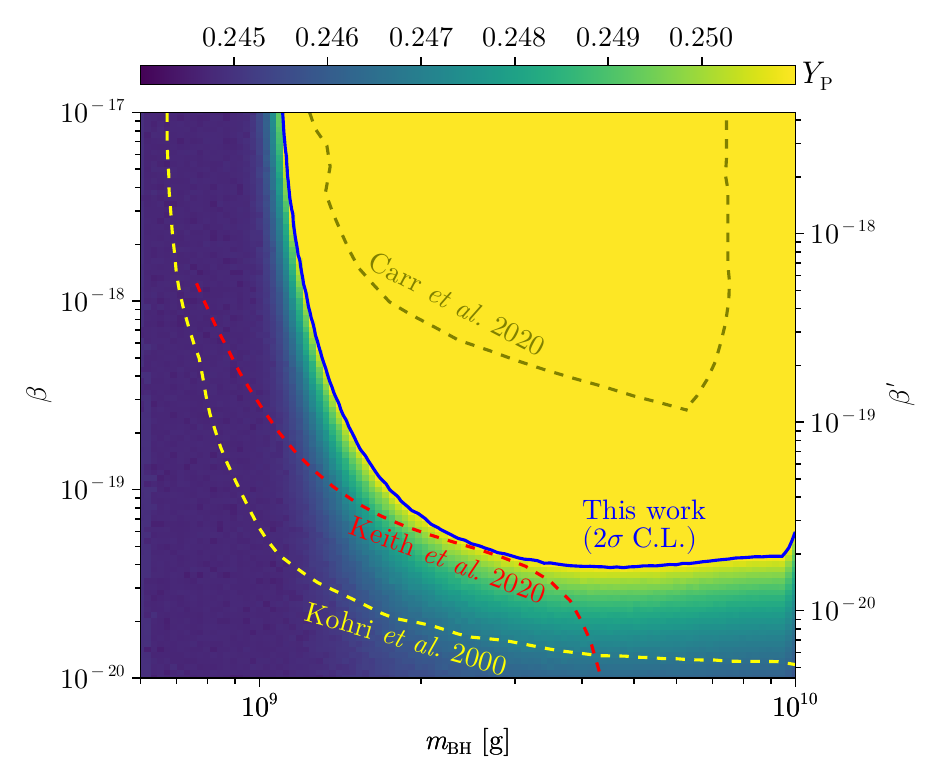}

\caption{The 2 $\sigma$ C.L. constraint on PBHs from the BBN observable $Y_{P}$
obtained in this work (solid line) compared with constraints obtained
in earlier studies (dashed lines), including Carr {\it et al.} 2020~\cite{Carr:2020gox},
 Kohri {\it et al.} 2000~\cite{Kohri:1999ex}, Keith {\it et
al.} 2020~\cite{Keith:2020jww}.  \label{fig:result}}
\end{figure}

In Fig.~\ref{fig:result}, we present our results for $Y_{P}$ in
the range of $\beta\in[10^{-20},\ 10^{-17}]$ and $m_{{\rm BH}}\in[6\times10^{8},\ 10^{10}]\ \text{g}$.
For the convenience of comparison with results in previous studies,
we also show the corresponding values of $\beta'$ on the right $y$-axis.
It is defined as
\begin{equation}
\beta'\equiv\gamma^{1/2}\left[\frac{g_{\star}\left(T_{i}\right)}{106.75}\right]^{-1/4}\left(\frac{h}{0.67}\right)^{-2}\beta\thinspace,\label{eq:-61}
\end{equation}
where $h\equiv H_{0}/(100\text{km}/\text{s}/\text{Mpc})$ is the
reduced Hubble expansion rate. We derive the 2 $\sigma$ C.L. constraint
(solid line) using the PDG recommended value $Y_{P}=0.245\pm0.003$
as the latest combined measurement~\cite{ParticleDataGroup:2024cfk},
i.e., the region above the line corresponds to $Y_{P}>0.251$. 

As is shown in Fig.~\ref{fig:result}, when $m_{{\rm BH}}$ is below
$10^{9}$ g, the resulting $Y_{P}$ exhibits no discernible differences
with respect to the standard value. This is expected from Eq.~\eqref{eq:-9},
which implies that PBHs lighter than $10^{9}$ g should have evaporated
before $T\approx1.8$ MeV. Above this temperature, even if the PBHs
caused a significant effect, it would be rapidly washed out by thermal
interactions, as demonstrated by  the red curve in Fig.~\ref{fig:sol}.
This contrasts with two previous results  by Kohri {\it et al.}
2000~\cite{Kohri:1999ex} and Keith {\it et al.} 2020~\cite{Keith:2020jww},
in which the BBN constraints extend below $10^{9}$ g, probably due
to less accurate treatments of PBH evaporation in these studies. Compare
with  Carr {\it et al.} 2020~\cite{Carr:2020gox}, our bound shares
similar mass thresholds but is overall more restrictive. This might
arise from some technical differences in the treatment of hadron emissivity
and different experimental values of $Y_{P}$ used in the analysis.

Regarding other light elements produced in BBN, we computed their
abundances using the nuclear reaction network from {\tt BBN-simple}~\cite{Meador-Woodruff:2024due}
and found that their sensitivities are much weaker than $Y_{P}$ if
the PBHs evaporate before BBN.  This has also been confirmed by Fig.~2
of Ref.~\cite{Kohri:1999ex}. Hence we do not include other elements
in our analysis.  We note here, however, that for heavier PBHs with
$m_{\mathrm{BH}}\gtrsim10^{10}$ g, other light elements may provide
more restrictive constraints. For instance, the abundance of deuterium
($\text{D}$) could be significantly affected by photo- and hadro-dissociation
of heavier elements such as $^{4}\text{He}$. Given  that the $^{4}\text{He}$
abundance is four orders of magnitude higher than the $\text{D}$
abundance after BBN, even a very small fraction of $^{4}\text{He}$
dissociated by radiations from PBHs would drastically increase the
$\text{D}$ abundance (e.g., 1\% of $^{4}\text{He}$ being dissociated
and converted to $\text{D}$ would increase the $\text{D}$ abundance
by two orders of magnitude). Therefore, for PBHs evaporating in the
post-BBN era, the constraint  from deuterium is much stronger than
that from helium~\cite{Kohri:1999ex,Carr:2020gox}. Since the post-BBN
analysis involves photo- and hadro-dissociation, which is very different
from the physics involved in this work, we leave it for future work
and plan to present the analysis in a companion paper on PBHs evaporating
after BBN. 

\section{Conclusion \label{sec:conclusion}}

PBHs evaporating before the onset of BBN can significantly alter its
successful predictions of the primordial light-element abundances.
In this work, we have investigated the impact of such PBHs on BBN
and derived the corresponding constraints.   We presented detailed
calculations of background evolution, hadronization of Hawking radiation,
meson-driven neutron--proton conversion, and the evolution of the
neutron-to-proton ratio. At each step, we compared the numerical
outputs of our code with analytical estimates and  found good agreements.
This enhances the transparency of our calculation and, together with
our publicly available code~\GitHub, ensures the reproducibility
of our results. 

Our main result is presented in Fig.~\ref{fig:result}. Compared
to  earlier studies, we observe significant discrepancies. Specifically,
according to our calculation, the PBH mass must exceed $10^{9}$ g
to cause observable effects on BBN, whereas the constraints in two
previous studies by Kohri {\it et al.} 2000~\cite{Kohri:1999ex}
and Keith {\it et al.} 2020~\cite{Keith:2020jww} extend below
this threshold, probably due to less accurate treatments of PBH evaporation.
Our result is closer to that of Carr {\it et al.} 2020~\cite{Carr:2020gox},
though our most restrictive bound occurs at $m_{{\rm BH}}\approx2\times10^{9}$
g, to be compared with $6\times10^{9}$ g from \cite{Carr:2020gox}.
 Overall, our constraint is weaker than those of \cite{Kohri:1999ex,Keith:2020jww},
but stronger than that of \cite{Carr:2020gox} except for $m_{{\rm BH}}/(10^{9}\ \text{g})\in[3.2,\ 7.4]$. 

Finally, we note that within the mass range shown in Fig.~\ref{fig:result},
PBHs evaporate prior to nucleosynthesis, causing no photo- and hadro-dissociation
effects. For PBHs with heavier masses, such effects become important
and require a dedicated analysis, which we plan to pursue in our upcoming
work.
\begin{acknowledgments}
We thank Bhupal Dev for helpful discussions on BBN physics, and also
Boting Zhou for his help in correcting a minor mistake in the calculation
of neutron-proton conversion rates.  This work is supported in part
by the National Natural Science Foundation of China under grant No.~12141501
and also by the CAS Project for Young Scientists in Basic Research
(YSBR-099). 
\end{acknowledgments}

\appendix

\section{Numerical details of the $\nu$-$\gamma$ temperature splitting \label{sec:T-split}}

After neutrino decoupling, the temperature of photons $T_{\gamma}$
starts to deviate from the temperature of neutrinos $T_{\nu}$, implying
that the SM thermal plasma splits into two sectors with different
temperatures. Both $T_{\gamma}$ and $T_{\nu}$ are important for
BBN calculations. Here we introduce the numerical details of how we
compute the temperature splitting between them. 

Assuming that neutrinos after decoupling are unaffected by the subsequent
$e^{\pm}$ annihilation (i.e., the very small amount of energy and
entropy injected from $e^{\pm}$ to neutrinos after the decoupling
is negligible), the neutrino temperature $T_{\nu}$ simply scales
as $a^{-1}$. Therefore, given a value of $T_{\gamma}$, the simplest
method to determine $T_{\nu}$ is 
\begin{equation}
T_{\nu}\approx T_{\gamma}\min\left\{ \left[\frac{g_{\star,s}(T_{\gamma})}{g_{\star,s}(T_{\text{\ensuremath{\nu}dec}})}\right]^{\frac{1}{3}},\ 1\right\} ,\label{eq:-56}
\end{equation}
 which is obtained by combining $T_{\nu}\propto a^{-1}$ and Eq.~\eqref{eq:-25}.
In this expression, $T_{\text{\ensuremath{\nu}dec}}$ denotes the
temperature of neutrino decoupling. Under the assumption that at neutrino
decoupling $e^{\pm}$ are still highly relativistic, the specific
value of $T_{\nu{\rm dec}}$ is unimportant here because this assumption
leads to $g_{\star,s}(T_{\text{\ensuremath{\nu}dec}})=2+4\times\frac{7}{8}+6\times\frac{7}{8}=10.75$.
 At higher temperatures when $g_{\star,s}>10.75$, neutrinos are
tightly coupled to photons, i.e., $T_{\nu}=T_{\gamma}$, which is
ensured by the ``$\min$'' function in Eq.~\eqref{eq:-56}. 

Equation~\eqref{eq:-56} can be used to calculate $T_{\nu}$ from
$T_{\gamma}$, provided that $g_{\star,s}(T_{\gamma})$ has been calculated.
If $g_{\star,s}(T_{\gamma})$ has not been determined, a relatively
simple approach to obtain it without solving the Boltzmann equations
governing neutrino decoupling is to solve the following equations:
\begin{align}
\left[2T_{\gamma}^{3}+\frac{7}{8}\cdot4\cdot{\cal I}\left(\frac{m_{e}}{T_{\gamma}}\right)T_{\gamma}^{3}+\frac{7}{8}\cdot6\cdot T_{\nu}^{3}\right]a^{3} & ={\rm constant}\thinspace,\label{eq:-57}\\
T_{\nu}a & ={\rm constant}\thinspace,\label{eq:-58}
\end{align}
where 
\begin{equation}
{\cal I}\left(x\right)\equiv\frac{90}{7\pi^{4}}\int_{0}^{\infty}\frac{\xi^{2}d\xi}{\exp\left(\sqrt{\xi^{2}+x^{2}}\right)+1}\left[\sqrt{\xi^{2}+x^{2}}+\frac{\xi^{2}}{3\sqrt{\xi^{2}+x^{2}}}\right].\label{eq:-59}
\end{equation}
Starting from neutrino decoupling ($T_{\gamma}=T_{\nu}=T_{\nu{\rm dec}}$)
and setting an initial value of $a$, we first use Eq.~\eqref{eq:-58}
to determine $T_{\nu}$ at any subsequent moment for a given \emph{a}.
Then with the known values of $T_{\nu}$ and $a$, we solve Eq.~\eqref{eq:-57}
to obtain $T_{\gamma}$. Eventually, we get two arrays of $T_{\nu}$
and $T_{\gamma}$ for a generated array of $a$. These three arrays
allow us to determine one from another conveniently via interpolation.

\section{The neutron-proton conversion rates\label{sec:conversion-rates}}

Considering a generic process $n+X\leftrightarrow p+Y$ with $X$
and $Y$ some generic species, we can calculate the following collision
terms:
\begin{align}
C_{n\to p} & \equiv\int d\Pi_{n}d\Pi_{X}d\Pi_{Y}d\Pi_{p}f_{n}f_{X}(1\pm f_{Y})(1-f_{p})(2\pi\delta)^{4}|\overline{{\cal M}}|^{2}\thinspace,\label{eq:-13}\\
C_{p\to n} & \equiv\int d\Pi_{p}d\Pi_{Y}d\Pi_{X}d\Pi_{n}f_{p}f_{Y}(1\pm f_{X})(1-f_{n})(2\pi\delta)^{4}|\overline{{\cal M}}|^{2}\thinspace,\label{eq:-14}
\end{align}
where $f_{i}$ ($i\in\{x,X,Y,p\}$) denotes the phase space distribution
function of particle $i$; $d\Pi_{n}\equiv\frac{g_{i}d^{3}p_{i}}{(2\pi)^{3}2E_{i}}$
with $p_{i}$, $E_{i}$, and $g_{i}$ the momentum, energy, and multiplicity
of $i$; $(2\pi\delta)^{4}$ is the four-momentum delta function responsible
for momentum conservation, and $|\overline{{\cal M}}|^{2}$ is the
squared matrix element of the process with the bar indicating spin-averaging
(over all initial and final states).  We shall note here that it
is straightforward to generalize $X$ and $Y$ to multiple species,
i.e., $X\to(X_{1},\ X_{2},\ \cdots,X_{i})$ and $Y\to(Y_{1},\ Y_{2},\ \cdots,Y_{j})$.
In particular, $n\leftrightarrow p+\overline{\nu_{e}}+e^{-}$ can
be included by setting $X\to\varnothing$ and $Y\to(\overline{\nu}_{e},\ e^{-})$. 

The conversion rates $\Gamma_{n\to p}$ and $\Gamma_{p\to n}$ appearing
in Eq.~\eqref{eq:Xn-ode} are  related to the above collision terms
as follows:
\begin{equation}
\Gamma_{n\to p}=\frac{C_{n\to p}}{n_{n}}\thinspace,\ \ \Gamma_{p\to n}=\frac{C_{p\to n}}{n_{p}}\thinspace.\label{eq:-10}
\end{equation}
Physically they can be interpreted as the probability of a neutron
or a proton being converted from one to the other per unit time.

Let us first consider neutron decay, $n\to p+\overline{\nu}_{e}+e^{-}$
(for simplicity, we will use the shorthand $\overline{\nu}_{e}\to\nu$
and $e^{-}\to e$ below), and focus on its contribution to $\Gamma_{n\to p}$.
Using the non-relativistic approximation of nucleons and neglecting
Pauli blocking factors, the contribution reads
\begin{align}
\Gamma_{n\to p}^{({\rm decay})} & \approx\frac{1}{2m_{n}}\int d\Pi_{\nu}d\Pi_{e}d\Pi_{p}(2\pi\delta)^{4}|\overline{{\cal M}}|^{2}\thinspace,\label{eq:-45}
\end{align}
where we have replaced $\int d\Pi_{n}f_{n}\to\frac{n_{n}}{2m_{n}}$,
as  justified by  the non-relativistic approximation. Equation~\eqref{eq:-45}
is exactly the decay rate of $n$. Here we shall clarify a subtlety
that may cause confusion between spin-summed and spin-averaged matrix
elements. When calculating a decay rate, one usually sums over the
spins of the final states and averages over the spins of the initial
states. In Eq.~\eqref{eq:-45}, however, we average over the spins
of both the initial and final, leading to a smaller matrix element
than in the former case. This reduction  is compensated by the multiplicity
factor $g_{i}$ in $d\Pi_{i}$. So the final result remains equivalent
to the conventional calculation  of a decay rate. 

The spin-averaged matrix element can be obtained from the effective
Lagrangian of weak interactions:
\begin{equation}
{\cal L}\supset\frac{G_{F}}{\sqrt{2}}\left[\overline{\psi}_{p}\gamma^{\mu}\left(g_{V}-g_{A}\gamma^{5}\right)\psi_{n}\right]\left[\overline{\psi}_{e}\gamma^{\mu}\left(1-\gamma^{5}\right)\psi_{\nu}\right]\label{eq:-47}
\end{equation}
with $G_{F}$ the Fermi constant and $\left(g_{V},\ g_{A}\right)\approx\left(1,\ 1.3\right)$.
 From the Lagrangian, the spin-averaged matrix element reads
\begin{align}
|\overline{{\cal M}}|_{n\to p+\nu+e}^{2}={} & \frac{1}{2^{4}}\cdot G_{F}^{2}\left[\left(g_{V}+g_{A}\right)^{2}\left(p_{n}\cdot p_{\nu}\right)\left(p_{p}\cdot p_{e}\right)+\left(g_{V}-g_{A}\right)^{2}\left(p_{n}\cdot p_{e}\right)\left(p_{p}\cdot p_{\nu}\right)\right.\nonumber \\
 & \hphantom{\frac{1}{2^{4}}\cdot G_{F}^{2}\quad}\left.{}+\left(g_{V}^{2}-g_{A}^{2}\right)m_{n}m_{p}\left(p_{e}\cdot p_{\nu}\right)\right]\,,
\end{align}
where we have added the factor of $1/2^{4}$ to explicitly account
for spin-averaging of all initial and final states. At the BBN temperatures
($T\sim\mathrm{MeV}$), the nucleons are effectively at rest, i.e.,
$p_{n}=(m_{n},\mathbf{0})$ and $p_{p}=(m_{p},\mathbf{0})$. Furthermore,
we may approximate $m_{n}\approx m_{p}$ since $Q\equiv m_{n}-m_{p}=1.3~\mathrm{MeV}\ll m_{p}\approx940~\mathrm{MeV}$.
Under this limit, the scalar products involving nucleons simplify
to
\begin{equation}
p_{n}\cdot p_{e}\approx p_{p}\cdot p_{e}\approx m_{p}E_{e}\quad\text{and}\quad p_{n}\cdot p_{\nu}\approx p_{p}\cdot p_{\nu}\approx m_{p}E_{\nu}\,,
\end{equation}
while the leptonic product remains 
\begin{equation}
p_{e}\cdot p_{\nu}=E_{e}E_{\nu}-\boldsymbol{p}_{e}\boldsymbol{\cdot}\boldsymbol{p}_{\nu}\,.
\end{equation}
Therefore, the spin-averaged matrix element becomes
\begin{equation}
|\overline{{\cal M}}|_{n\to p+\nu+e}^{2}\approx\frac{1}{2^{4}}\cdot G_{F}^{2}m_{p}^{2}\left[\left(g_{V}^{2}+3g_{A}^{2}\right)E_{e}E_{\nu}+\left(g_{V}^{2}-g_{A}^{2}\right)\left(\boldsymbol{p}_{e}\boldsymbol{\cdot}\boldsymbol{p}_{\nu}\right)\right]\,.\label{eq:-46}
\end{equation}
Note that the term proportional to $\boldsymbol{p}_{e}\boldsymbol{\cdot}\boldsymbol{p}_{\nu}=p_{e}p_{\nu}\cos\theta$
does not contribute to the total decay rate when $\cos\theta$ is
integrated out. A more careful treatment could include a small correction
from  the CKM mixing---see, e.g., Eqs.~(5.139-5.140) in Ref.~\cite{Giunti:2007ry}.

It is worth mentioning that \emph{crossing symmetry} dictates that
$\nu+n\to p+e$ and $e+n\to p+\nu$ have the same spin-averaged matrix
element (the minus sign arising from crossing a fermionic leg is canceled
by $p_{\nu}\to-p_{\nu}$ or $p_{e}\to-p_{e}$)~\cite{crossing-symmetry,Bellazzini:2016xrt}:
\begin{equation}
|\overline{{\cal M}}|_{\nu+n\to p+e}^{2}=|\overline{{\cal M}}|_{e+n\to p+\nu}^{2}=|\overline{{\cal M}}|_{n\to p+\nu+e}^{2}\thinspace.\label{eq:-48}
\end{equation}
Although these matrix elements are formally identical, the scalar
products of momenta (e.g., $p_{n,p}\cdot p_{e}$, $p_{n,p}\cdot p_{\nu}$,
and $p_{\nu}\cdot p_{e}$) in each of them depends on different kinematics. 

\begin{figure}
\centering

\includegraphics[width=0.7\textwidth]{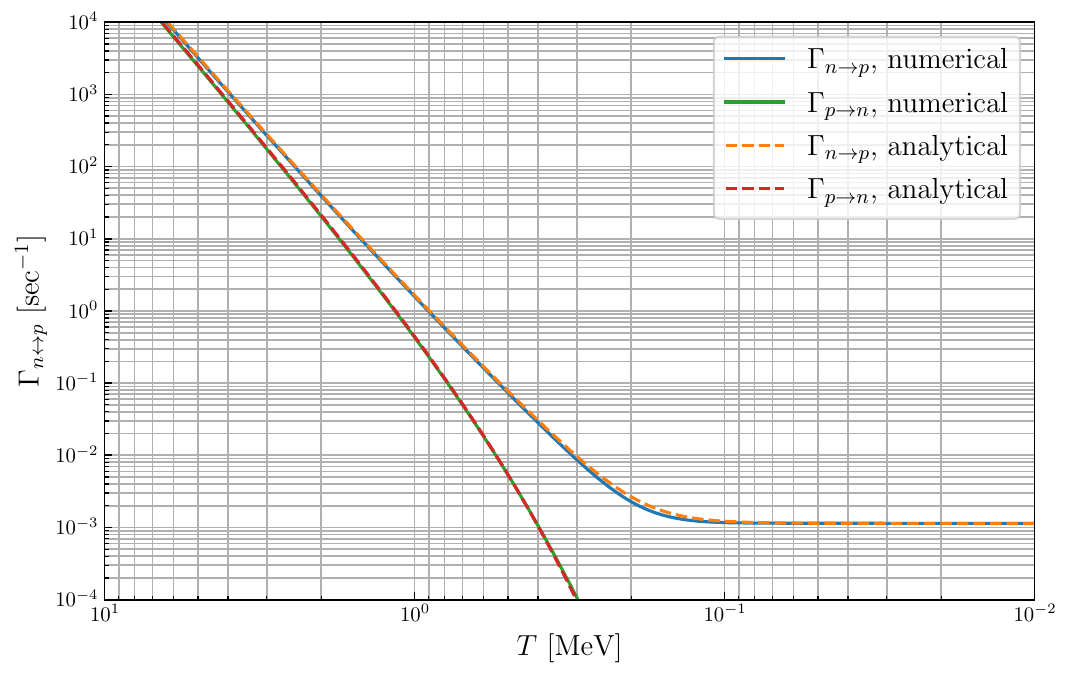}

\caption{\label{fig:compare-rate} Comparison of the analytical neutron-proton
conversion rates {[}Eqs.~\eqref{eq:-53} and \eqref{eq:-55}{]} with
the numerical ones from Ref.~\cite{Meador-Woodruff:2024due}.}
\end{figure}

Substituting Eq.~\eqref{eq:-46} into Eq.~\eqref{eq:-45}, we obtain
\begin{align}
\Gamma_{n\to p}^{({\rm decay})} & \approx\frac{g_{p}}{4m_{n}m_{p}}\int d\Pi_{\nu}d\Pi_{e}2\pi\delta(Q-E_{\nu}-E_{e})|\overline{{\cal M}}|^{2}\nonumber \\
 & \approx\frac{g_{p}g_{\nu}g_{e}G_{F}^{2}}{2}\int\frac{4\pi p_{\nu}^{2}dp_{\nu}}{(2\pi)^{3}2E_{\nu}}\frac{2\pi p_{e}^{2}dp_{e}dc_{\theta}}{(2\pi)^{3}2E_{e}}2\pi\delta(Q-E_{\nu}-E_{e})\nonumber \\
 & \hphantom{\approx\frac{g_{p}g_{\nu}g_{e}G_{F}^{2}}{2}\int\frac{4\pi p_{\nu}^{2}dp_{\nu}}{(2\pi)^{3}2E_{\nu}}}{}\times\left[\left(g_{V}^{2}+3g_{A}^{2}\right)E_{\nu}E_{e}+\left(g_{V}^{2}-g_{A}^{2}\right)p_{e}p_{\nu}c_{\theta}\right]\nonumber \\
 & \approx\frac{g_{p}g_{\nu}g_{e}G_{F}^{2}g_{VA}^{2}}{2}\int\frac{4\pi p_{\nu}^{2}dp_{\nu}}{(2\pi)^{3}2E_{\nu}}\frac{4\pi p_{e}^{2}dp_{e}}{(2\pi)^{3}2E_{e}}2\pi\delta(Q-E_{\nu}-E_{e})E_{\nu}E_{e}\nonumber \\
 & \approx\frac{g_{p}g_{\nu}g_{e}G_{F}^{2}g_{VA}^{2}}{16\pi^{3}}\int_{0}^{p_{e}^{\max}}dp_{e}(Q-E_{e})^{2}p_{e}^{2}\nonumber \\
 & \approx\frac{g_{p}g_{\nu}g_{e}G_{F}^{2}g_{VA}^{2}m_{e}^{5}}{16\pi^{3}}\lambda_{0}\thinspace,\label{eq:-50}
\end{align}
where $c_{\theta}\equiv(\mathbf{p}_{\nu}\cdot\mathbf{p}_{e})/(p_{\nu}p_{e})$,
$g_{VA}^{2}\equiv g_{V}^{2}+3g_{A}^{2}$, $p_{e}^{\max}=\sqrt{Q^{2}-m_{e}^{2}}$,
and 
\begin{equation}
\lambda_{0}\equiv\int_{0}^{p_{e}^{\max}}dp_{e}\frac{(Q-E_{e})^{2}p_{e}^{2}}{m_{e}^{5}}\approx1.636\thinspace.\label{eq:-49}
\end{equation}
Using the value in Eq.~\eqref{eq:-49} and $\left(g_{V},\ g_{A}\right)\approx\left(1,\ 1.3\right)$,
we obtain $1/\Gamma_{n\to p}^{({\rm decay})}\approx8.7\times10^{2}\ \text{s}$,
which is close to the neutron lifetime $\tau_{n}$. The difference
is caused by various next-to-leading-order corrections including the
Coulomb attraction between $p$ and $e$. 

By slightly modifying the integral of Eq.~\eqref{eq:-50}, we can
straightforwardly obtain the conversation rate of $\nu+n\to p+e$:
\begin{align}
\Gamma_{n\to p}^{(\nu)} & \approx\frac{g_{p}}{4m_{n}m_{p}}\int d\Pi_{\nu}d\Pi_{e}f_{\nu}2\pi\delta(Q+E_{\nu}-E_{e})|\overline{{\cal M}}|^{2}\nonumber \\
 & \approx\frac{g_{p}g_{\nu}g_{e}G_{F}^{2}g_{VA}^{2}}{16\pi^{3}}\int_{Q}^{\infty}dE_{e}(E_{e}-Q)^{2}p_{e}E_{e}f_{\nu}\nonumber \\
 & \approx\frac{g_{p}g_{\nu}g_{e}G_{F}^{2}g_{VA}^{2}}{16\pi^{3}}\cdot2T^{3}\left(Q^{2}+6QT+12T^{2}\right),\label{eq:-52}
\end{align}
where in the second step we have used the relativistic approximation
$p_{e}\approx E_{e}$ and the Boltzmann approximation $f_{\nu}\approx e^{-E_{\nu}/T}=e^{-(E_{e}-Q)/T}$.

The calculation for $e+n\to p+\nu$ is similar (except that $(E_{e}+Q)^{2}\to(E_{e}-Q)^{2}$,
$\int_{Q}^{\infty}dE_{e}\to\int_{m_{e}}^{\infty}dE_{e}$, and $f_{\nu}\to f_{e}$)
and leads to 
\begin{equation}
\Gamma_{n\to p}^{(e)}\approx\Gamma_{n\to p}^{(\nu)}\thinspace.\label{eq:-51}
\end{equation}

Combining Eqs.~\eqref{eq:-49}, \eqref{eq:-52}, and \eqref{eq:-51},
we obtain the SM contribution to the $n\to p$ conversion rate:
\begin{equation}
\Gamma_{n\to p}^{({\rm SM})}\approx\frac{1}{\tau_{n}}\left[1+\frac{4Q^{5}}{m_{e}^{5}\lambda_{0}}\cdot\frac{12+6x+x^{2}}{x^{5}}\right]\label{eq:-53}
\end{equation}
with
\begin{equation}
x\equiv\frac{Q}{T}\thinspace,\ \ \text{ \ensuremath{\frac{4Q^{5}}{m_{e}^{5}\lambda_{0}}}\ensuremath{\ensuremath{\approx}253.9}}\thinspace.\label{eq:-54}
\end{equation}

The calculation of $\Gamma_{p\to n}^{({\rm SM})}$ is very similar
and leads to
\begin{equation}
\Gamma_{p\to n}^{({\rm SM})}\approx e^{-x}\Gamma_{n\to p}^{({\rm SM})}\thinspace.\label{eq:-55}
\end{equation}

In deriving Eqs.~\eqref{eq:-53} and \eqref{eq:-55}, we adopted the
Boltzmann approximation (neglecting Pauli blocking factors) and, at
certain stages, the relativistic approximation $p_{e}\approx E_{e}$.
A more rigorous treatment without these approximations would require
numerical integration\,---\,see, e.g., Eq.~(28) of Ref.~\cite{Meador-Woodruff:2024due}.
In Fig.~\ref{fig:compare-rate}, we compare the analytical results
given by Eqs.~\eqref{eq:-53} and \eqref{eq:-55} with the numerical
calculation of Ref.~\cite{Meador-Woodruff:2024due}. The close agreement
between the two implies that the analytical expressions provide adequate
accuracy for most applications, with deviations becoming relevant
only in calculations demanding the highest precision, such as a dedicated
BBN analysis for the SM.

\section{PYTHIA implementation details\label{sec:Pythia}}

When quarks ($q$) and gluons ($g$) are emitted by PBHs, they undergo
hadronization and produce mesons and other color singlets.  The treatment
 of hadronization requires dedicated packages like {\tt PYTHIA}.
In this appendix, we present the details of our {\tt PYTHIA} implementation. 

Our approach is  similar to how {\tt Blackhawk} invokes {\tt PYTHIA}:
the SM processes $Z\to q\overline{q}$ and $H\to gg$ are employed
to generate quarks and gluons, which then hadronize according to {\tt
PYTHIA}'s built-in hadronization module. To obtain a clean output,
we turn off all decay channels of $Z$ and $H$ except for the one
being investigated. For on-shell $Z$ and $H$, the produced quarks
and gluons have energies $E_{q}=m_{Z}/2$ and $E_{g}=m_{H}/2$ where
$m_{Z}$ and $m_{H}$ are the masses of $Z$ and $H$. To compute
the hadronization of quarks and gluons with higher or lower energies,
we reset the masses to $2E_{q}$ or $2E_{g}$ in {\tt PYTHIA}. 

In {\tt Blackhawk}'s {\tt PYTHIA} scripts, $Z$ and $H$ are produced
via $e^{+}e^{-}$ collisions. In processes $e^{+}e^{-}\to Z\to q\overline{q}$
and $e^{+}e^{-}\to H\to g\overline{g}$, $Z$ and $H$ can be off-shell
particles. This also allows one to obtain the hadronization of quarks
and gluons at varying energies. In this off-shell approach, one sets
the center-of-mass energy $E_{{\rm CM}}$ of the electron and positron
beams. Then the energy of quarks or gluons is $E_{{\rm CM}}/2$. We
have compared the outputs of both approaches (on-shell  versus off-shell)
and find that, if the off-shell approach incorporates a few adjustments
(see discussions later), they generate the same result. 

Our on-shell approach is implemented by the following {\tt PYTHIA}
settings.\begin{center}
\fbox{\begin{minipage}[t]{0.7\textwidth}
{\tt

pythia.readString(mother+\textquotedbl :m0 = \textquotedbl +ECM); 

pythia.readString(mother+\textquotedbl :onMode = off\textquotedbl ); 

pythia.readString(mother+\textquotedbl :onIfAny = \textquotedbl +daughter); 

pythia.readString(\textquotedbl PartonLevel:FSR = on\textquotedbl ); 

pythia.readString(\textquotedbl HadronLevel:all = on\textquotedbl );

}\end{minipage}
}\end{center}

Here {\tt mother} and {\tt daughter} are the particle IDs (in
PDG convention) of the mother and daughter particles of the decay
processes, respectively. For instance, for  $Z\to u\overline{u}$,
we set {\tt mother = \textquotedbl 23\textquotedbl} and {\tt
daughter = \textquotedbl 2\textquotedbl}. {\tt ECM} is the center-of-mass
energy of the quark or gluon pair. For instance, to study the hadronization
of a 1 TeV quark, we set {\tt ECM = \textquotedbl 2000.\textquotedbl}. 

Then in the event generation {\tt for} loop, we inject the following
code:\begin{center}
\fbox{\begin{minipage}[t]{0.9\textwidth}
{\tt

pythia.event.reset(); 

pythia.event.append(mother, 1, 0, 0, 0.0, 0.0, 0.0, ECM, ECM);

}\end{minipage}
}\end{center}

\begin{figure}
\centering

\includegraphics[width=0.6\textwidth]{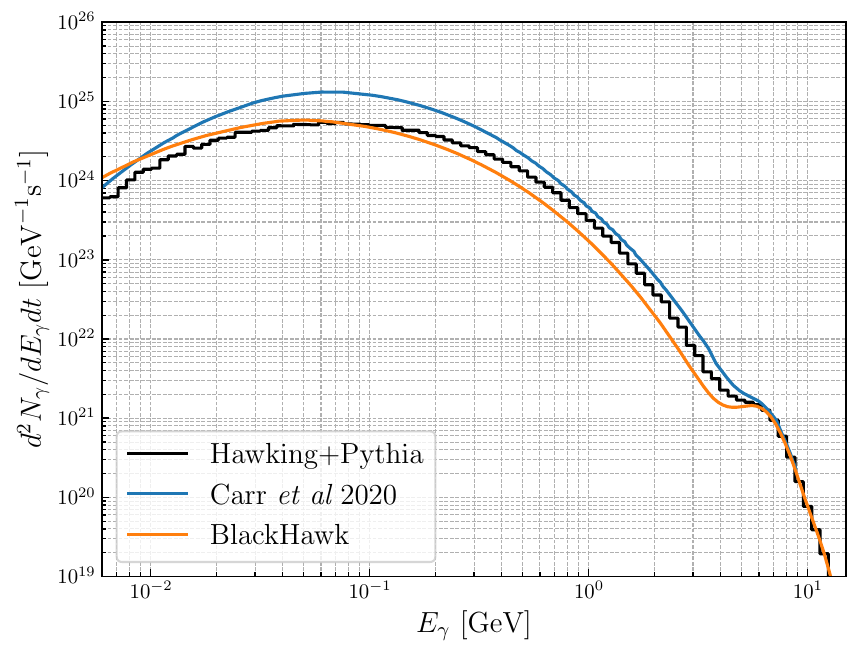}

\caption{The photon spectrum of a PBH with $T_{{\rm BH}}=1$ GeV, including
both primary and secondary components. The black line is generated
using Hawking radiation followed by hadronization simulated in {\tt
PYTHIA}. The orange line is a straightforward output of {\tt BlackHawk},
and the blue line is taken from Ref.~\cite{Carr:2020gox}. \label{fig:photon-1GeV}}
\end{figure}

Alternatively, one could also choose the off-shell approach, as already
implemented by {\tt Blackhawk}. In this approach, we shall raise
a few issues that require extra settings in {\tt PYTHIA}, otherwise
they can significantly affect the results of hadronization at low
energies. At high energies (well above all implicit energy cuts in
{\tt PYTHIA}), their influence is negligible. The first issue is
that one should turn of the lepton PDF via {\tt\textquotedbl PDF:lepton
= off\textquotedbl}. This is missing in the {\tt Blackhawk} implementation,
rendering the produced quarks and gluons slightly less energetic than
$E_{{\rm CM}}/2$ due to photon emission in $e^{+}e^{-}$ collisions.
Second, the phase space cuts need to be set properly. By default,
{\tt PYTHIA} sets the minimum invariant mass at 4 GeV via {\tt\textquotedbl PhaseSpace:mHatMin
= 4.0\textquotedbl}. We recommend setting {\tt\textquotedbl PhaseSpace:mHatMin
= 0.\textquotedbl} together with {\tt\textquotedbl PhaseSpace:pTHatMin
= 0.\textquotedbl} when {\tt PYTHIA} is used to compute hadronization
of quarks and gluons below a few GeV. The third issue is related to
the mass setting of $Z$ and $H$. When $Z$ and $H$ are produced
from $e^{+}e^{-}$ collisions with $E_{{\rm CM}}$ below a threshold
(the default value of this threshold in {\tt PYTHIA} is 10.0 GeV
for $Z$ and $50.0$ GeV for Higgs), the event will be abandoned,
disregarding that they serve as off-shell mother particles for quark
and gluon production. To avoid this cut, we suggest setting {\tt\textquotedbl 23:mMin
= 0.0\textquotedbl} and {\tt\textquotedbl 25:mMin = 0.0\textquotedbl}
for $Z$ and $H$ respectively. In addition, we notice that for $e^{+}e^{-}\to H$,
one also needs to set {\tt\textquotedbl 25:m0 = X\textquotedbl}
with {\tt X} slightly below $E_{{\rm CM}}$ to pass the cut. Finally,
let us mention that the {\tt PYTHIA} setting {\tt\textquotedbl WeakSingleBoson:ffbar2ffbar(s:gmZ)
= on\textquotedbl} used in {\tt Blackhawk} can be replaced with
{\tt\textquotedbl WeakSingleBoson:ffbar2gmZ = on\textquotedbl}
and {\tt\textquotedbl WeakZ0:gmZmode = 2\textquotedbl} to avoid
potential interference between $\gamma$ and $Z$, though in practice
we have not noticed any significant differences in the results. 

Obviously, the off-shell approach requires more careful adjustments
due to various hidden cuts set by default in {\tt PYTHIA}. Therefore,
if {\tt PYTHIA} is used independently of {\tt Blackhawk} to compute
the hadronization of Hawking radiation, we recommend the on-shell
approach.

With the {\tt PYTHIA} implementation explained in details, it is
straightforward to obtain the numbers and energy distributions of
various mesons produced from the hadronization of quarks and gluons.
The subsequent decays of these mesons are also automatically simulated
in {\tt PYTHIA}. This allows us to readily obtain the secondary
spectrum of, for example, photons, which at low energies are abundantly
generated from mesons (mostly from $\pi^{0}\to2\gamma$). In Fig.~\ref{fig:photon-1GeV},
we present the photon spectrum of a PBH obtained in this way (the
black line). Here we set the PBH temperature at $T_{{\rm BH}}=1$
GeV. Both primary and secondary components are included. For comparison,
we also present results  from {\tt BlackHawk} (orange line) and
from Ref.~\cite{Carr:2020gox} (blue line). One can see that while
the primary parts (dominant at $E_{\gamma}\gtrsim5$ GeV) are in very
good agreement with each other, the secondary parts differ by a factor
of a few. Since the secondary photons are predominantly produced from
hadronic processes,  the comparison implies that the details of hadronization
implementation which may vary in the literature are of significant
importance. 

\bibliographystyle{JHEP}
\bibliography{ref}

\end{document}